\documentclass{aa}
\usepackage{graphicx}
\usepackage{times}

\begin{document}

\def\grad{\hbox{$^\circ$}}
\def\bsec{\hbox{$^{\prime\prime}$}}
\def\std{\hbox{$^{\rm h}$}}
\def\min{\hbox{$^{\rm m}$}}
\def\sec{\hbox{$^{\rm s}$}}
\def\mag{\mbox{ mag}}

\thesaurus{(08.08.1;         
            08.11.1;         
            11.19.4;         
            11.13.1)         
            }

\title{Studies of Binary Star Cluster Candidates in the Bar of the LMC. I: SL\,353 \& SL\,349
\thanks{Based on observations taken at the European Southern Observatory, La
  Silla, Chile.}}

\author{Andrea Dieball\inst{1}, Eva K. Grebel\thanks{Hubble fellow}\inst{2},
  Christian Theis\inst{3}}

\institute{Sternwarte der Universit\"at Bonn, Auf dem H\"ugel 71, 
           D--53121 Bonn, F.R. Germany \and
           University of Washington, Department of Astronomy, Box 351580, 
           Seattle, WA 98195-1580, USA \and
           Institut f\"ur Theoretische Physik und Astrophysik der
           Universit\"at Kiel, D-24098 Kiel, F.R. Germany} 

\offprints{Andrea Dieball, adieball@astro.uni-bonn.de}

\date{Received 22 December 1999, accepted 27 March 2000}

\maketitle

\markboth{A. Dieball et al.: Binary Clusters in the LMC Bar. I}{A. Dieball et
  al.: Binary Clusters in the LMC Bar. I} 

\begin{abstract}

We present a detailed study of the cluster pair SL\,353 \& SL\,349. This
candidate binary cluster is located at the northwestern rim of the LMC
bar. Based on photometric data we find that both clusters are coeval with an
age of 550$\pm$100 Myr. We use the Ca\,II triplet in the spectra of individual
red giants to derive radial velocities. Both components of the binary cluster
candidate show very similar mean velocities ($\approx274\pm10$
km\,$\mbox{s}^{-1}$ for SL\,349 and $\approx279\pm4$ km\,$\mbox{s}^{-1}$ for
SL\,353) while the field stars show lower velocities ($\approx240\pm19$
km\,$\mbox{s}^{-1}$). These findings suggest a common origin of the two
clusters from the same GMC. In this sense the cluster pair may constitute a
true binary cluster. We furthermore investigate the stellar densities in and
around the star clusters and compare them with isopleths created from
artificial, interacting as well as non-interacting star
clusters. Gravitational interaction leads to a distortion which can also be
found in the observed pair.  


\end{abstract}

\section{Introduction}

The Magellanic Clouds offer the unique possibility to study star clusters in
general and binary clusters in particular. These two companion galaxies are 
close enough to resolve single stars, but distant enough to make the detection
of close pairs of star clusters an easy task. About a decade ago, Bhatia \&
Hatzidimitriou (\cite{bh}), Hatzidimitriou \& Bhatia (\cite{hb}), and Bhatia
et al. (\cite{brht}) surveyed the Magellanic Clouds in order to catalogue 
binary cluster candidates. To qualify as a binary cluster candidate, the
maximum projected centre-to-centre separation of the components of a pair was
chosen to be $\approx 1\farcm3$, which corresponds to $\approx$ 19 pc in the
LMC if a distance modulus of 18.5 mag is adopted. A binary cluster with larger
separation may become detached by the external tidal forces while shorter
separations may lead to mergers (Sugimoto \& Makino \cite{sm}, Bhatia
\cite{bhatia}). The number of chance-pairs of objects uniformly distributed in
space can be estimated adopting a formula presented by Page (1975): Roughly
half of the pairs found may be explained by mere chance line-up. This suggests
that at least a subset of them must be true binary clusters, i.e., clusters
that are formed together and/or may interact or even be gravitationally
bound. 
       
Depending on their masses and separations binary clusters will eventually
merge or become detached. At one stage during the merger process the former
binary cluster could have one single but elliptical core (Bhatia \&
McGillivray \cite{bm}). Recently de\,Oliveira et al. (\cite{obd}) performed
numerical simulations of star cluster encounters which could represent a
possible scenario to explain the ellipticities found in several star clusters
in the Magellanic Clouds.    

Star clusters form in giant molecular clouds (GMCs), but the details of cluster
formation are not yet understood (Elme\-green et al. \cite{eepz}). Fujimoto \&
Kumai (\cite{fk}) suggest that binary or multiple star clusters form through
strong, oblique collisions between massive gas clouds in high-velocity random
motion, resulting in compressed sub-clouds revolving around each other. Since
the components of a cluster pair formed together, they should be coeval or at
least have a small age difference compatible with cluster formation time
scales. Binary clusters are expected to form more easily in galaxies like the
Magellanic Clouds, whereas in the Milky Way the required large-scale
high-velocity random motions are lacking.  
Indeed only few binary open cluster candidates and no globular cluster pairs
are known in our Galaxy (see for example Subramaniam \& Sagar
\cite{ss}). However, binary globular clusters are not expected to survive the
gravitational forces of the Milky Way (Surdin \cite{surdin}).    

In case of tidal capture two clusters would be gravitationally bound, but age
differences are likely. Encounters of clusters can be traced using isodensity
maps (de\,Oliveira et al. \cite{dodb}, Leon et al. \cite{lbv}). Though
van\,den\,Bergh (\cite{vdbergh}) suggests that tidal capture becomes more
probable in dwarf galaxies like the Magellanic Clouds with small velocity
dispersion of the cluster system, Vallenari et al. (\cite{vbc}) estimated a
cluster encounter rate of $dN/dt\sim1\cdot(10^{9} \mbox{yr})^{-1}$. This makes
tidal capture of young clusters very unlikely. 
      
The formation of low-mass star clusters tends to proceed hierarchically in
large molecular complexes over several $10^7$ years (e.g., Efremov \&
Elmegreen \cite{ee}). Leon et al. (\cite{lbv}) suggest that in these groups
the cluster encounter rate is higher and thus tidal capture is more likely:
Binary clusters are not born together as a pair with similar ages but are
formed later through gravitational capture. An observational test of this
scenario would require the detection of evidence of tidal interactions between
clusters, whose age differences need to be compatible with the survival times
of GMCs. 

Another binary cluster formation scenario is introduced by Theis
(\cite{theis}) and Ehlerov\'a et al. (\cite{epth}): Exploding supernovae close
to the centre of a GMC sweep up the outer cloud material
within a few Myrs and accumulate it in the shell. The large amount of matter
makes the shell prone to gravitational fragmentation and may eventually lead
to the formation of many open cluster-like associations (Theis et
al. \cite{teph}, Ehlerov\'a et al. \cite{epth}). In case of a dense ambient
medium outside the cloud or a very massive original molecular cloud the shell
can be strongly decelerated resulting in a gravitationally bound system of
fragments or stars which can recollapse. A galactic tidal field acting on this
recollapsing shell can split it into two or more large clusters, thus forming
coeval twin globulars (Theis \cite{theis}). These clusters may stay together
for a long time, though they are gravitationally unbound. The evolution of
their spatial separation mainly depends on the shape of the shock front at the
time of fragmentation.  

We are studying binary cluster candidates in the Magellanic Clouds to
investigate whether the cluster pairs may be of common origin and if they show
evidence for interaction. While it is so far impossible to measure true,
deprojected distances between apparent binary clusters, an analysis of their
properties can give clues to a possible common origin. 

The binary cluster candidate SL\,353 (or BRHT\,33b, see Bhatia et
al. \cite{brht}) and SL\,349 (BRHT\,33a) is located in the outer western part
of the LMC bar. Based on integrated colours, Bica et al. (\cite{bcdsp})
suggest that SL\,353 \& SL\,349 are coeval clusters of  SWB type V (Searle et
al. \cite{swb}) which is in agreement with the findings of Vallenari et
al. (\cite{vbc}). Leon et al. (\cite{lbv}) suggest that the two clusters may
even be physically connected. Based on ages for a large sample of star
clusters, derived on the base of integrated colours, Bica et
al. (\cite{bcdsp}) propose an age gradient of the LMC bar. Younger clusters
are predominantly found in the eastern part, while older clusters of SWB type
III and higher are concentrated to its western end. 

This paper is organized as follows. In Sect. \ref{phot} we describe the
photometric data in general. Stellar density maps are presented in
Sect. \ref{stardens}. The following section describes the colour magnitude
diagrams (CMD) for the components of the cluster pair. Ages for each cluster
are derived and compared with previous studies. The spectroscopic data are
described in Sect. \ref{spectroscopy}, and radial velocities are derived in
Sect. \ref{radvel}. A comparison of the observed isopleths with simulated ones
for interacting and non-interacting star clusters is given in
Sect. \ref{nbody}. In the last Sect. \ref{summary} we summarize and discuss
the results.   

\section{Photometric observations and data reduction}
\label{phot}

The data were obtained on March 22, 1994, with EFOSC 2 
at the ESO/MPI 2.2 m telescope at La Silla. A $1024 \times 1024$ coated
Thomson 31156 chip (ESO \#19) was used with a pixel scale of $0\farcs 34$
resulting in a field of view of $5\farcm8 \times 5\farcm8$.  These data were
obtained with the 
Gunn $g$ (which is similar to Washington $M$) and Gunn $i$ (corresponding to
Washington $T2$) filters used at the 2.2 m telescope. An observing log is
given in Table \ref{obs}. 
During standard image reduction with MIDAS we noticed flatfielding problems
near the edges of the exposures. Thus, each image was cut out resulting in a
field of view of $4\farcm3 \times 3\farcm5$.  
An image of the binary cluster candidate is shown in Fig. \ref{sl353ps}.

Profile fitting photometry was carried out with DAOPHOT~II (Stetson
\cite{stetson}) running under MIDAS. 

The photometry was transformed using the standard fields around SA\,110 and 
SA\,98 (Geisler \cite{geisler}) observed in the same night.

We applied the following transformation relations:
\begin{eqnarray*}
&&g-G = z_{G} + a_{G} \cdot X + c_{G} \cdot (G-I)\\
&&i-I = z_{I} + a_{I} \cdot X + c_{I} \cdot (G-I),
\end{eqnarray*}
where $X$ is the mean airmass during observation, capital letters
represent standard magnitudes and colours, and lower-case
letters denote instrumental magnitudes after normalizing to
an exposure time of 1 sec. The resulting colour terms $c_{i}$, zero
points $z_{i}$ and atmospheric extinction coefficients $a_{i}$ are listed in
table \ref{transf}.

\begin{table}
\caption[]{\label{obs}Observing log}
\begin{tabular}{crc}
\hline
Filter   &\multicolumn{1}{c}{Exp.time} [s] & Seeing [\arcsec] \\\hline
Gunn $g$ & 600                        & 1.6 \\
Gunn $g$ & 60                         & 1.4 \\
Gunn $g$ & 40                         & 1.4 \\
Gunn $i$ & 240                        & 1.1 \\
Gunn $i$ & 20                         & 1.2 \\ \hline
\hline
\end{tabular}
\end{table}

\begin{table}
\caption[]{\label{transf}Transformation coefficients}
\begin{tabular}{lrcc}
\hline
Filter   & \multicolumn{1}{c}{$c_{i}$} & $z_{i}$ &  $a_{i}$\\
         &         & [mag]   & [mag]   \\
\hline
Gunn $g$ & $-0.060\pm0.006$ & $1.685\pm0.030$ & $0.206\pm0.019$ \\
Gunn $i$ & $ 0.051\pm0.003$ & $2.439\pm0.016$ & $0.081\pm0.010$ \\ \hline
\end{tabular}
\end{table} 

\section{Stellar density maps}
\label{stardens}

\begin{figure}
\centerline{
\includegraphics[width=\hsize]{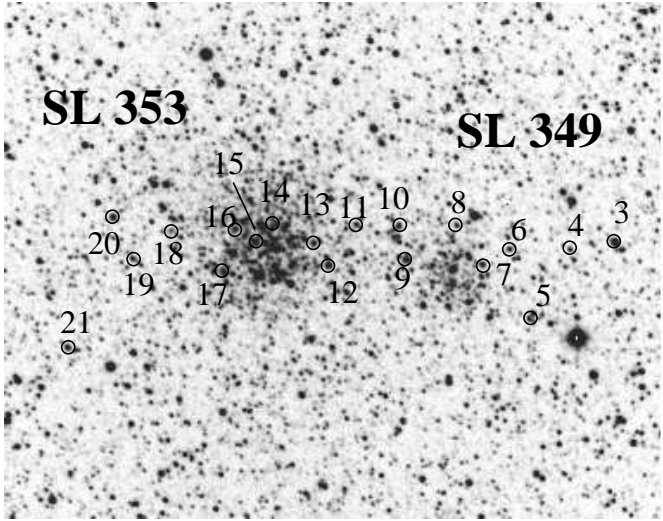}
}
\caption[]{\label{sl353ps} Gunn $i$-image of the cluster pair SL\,353 \&
  SL\,349. North is up and east to the left. The field of view of this image
  is $4\farcm3\times3\farcm5$. These clusters are located at the outer
  northwestern rim of the LMC bar. For the stars marked with circles also
  spectroscopic data were obtained (see Sect. \ref{spectroscopy})} 
\end{figure}

\begin{figure}
\centerline{
\includegraphics[width=\hsize]{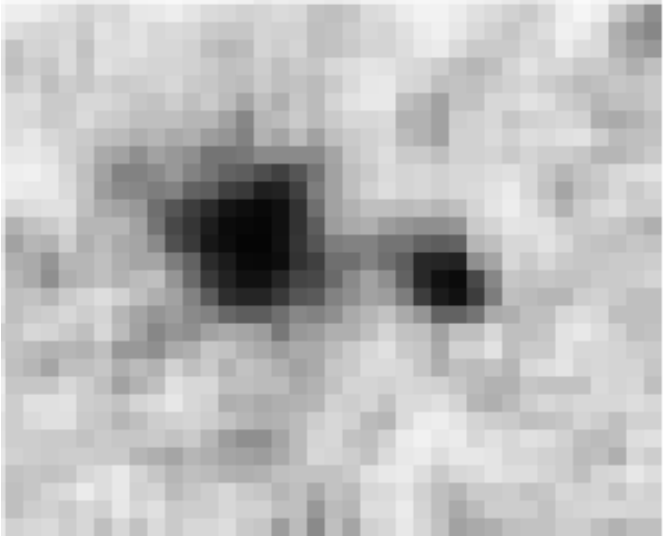}
}
\caption[]{\label{sl353dens} Star density map of SL\,353 \& SL\,349. The star
  density between the cluster pair is enhanced. Note the distortion
  of SL\,349 towards the bigger cluster SL\,353. Similar distortions are seen
  around the more massive cluster}
\end{figure}

The cluster pair SL\,353 \& SL\,349 (see Fig. \ref{sl353ps}) is located in the
inner northwestern part of the LMC at the outer rim of the bar. The projected
distance between the clusters' centres is $1\farcm24$. Assuming a distance
modulus of $m-M = 18.5$ mag (Westerlund \cite{westerlund}) this corresponds to
18.1 pc. According to 
Sugimoto \& Makino (\cite{sm}) this is
close to the maximum separation at which a binary  cluster is still stable and
will not be detached by the outer tidal forces of the LMC.

We investigate the stellar density distribution in and around the clusters by 
counting the number of stars in square cells of 20 pixels length
(corresponding to $6\farcs8$ or 1.7 pc). 
To make density structures and thus possible signs of interaction better
visible we applied a $3\times 3$ average filter for image smoothing. 
The procedure is described in  more detail in Dieball \& Grebel (\cite{dg}). 

The stellar density map is presented in Fig. \ref{sl353dens}. The star density
between the two clusters  is enhanced compared to the surrounding field
density. The enhancement is about 3 $\sigma$ above the field background and
thus significant at a $\approx99$~\%-level. In addition, the smaller cluster
SL\,349 seems to be warped towards SL\,353. Also the density profile of
SL\,353 shows distortions: we see smaller arcs of enhanced density reaching to
the north and south, and also to the northeast. Whether these features might
be tidal tails due to an interaction with the LMC tidal field or even due to
an interaction with the smaller companion SL\,349, as suggested by Leon et
al. (\cite{lbv}), is hard to decide.  

\section{Colour Magnitude Diagrams and Isochrone fitting}
\label{ages}

\begin{figure*}
\centerline{
\includegraphics[width=8.8cm]{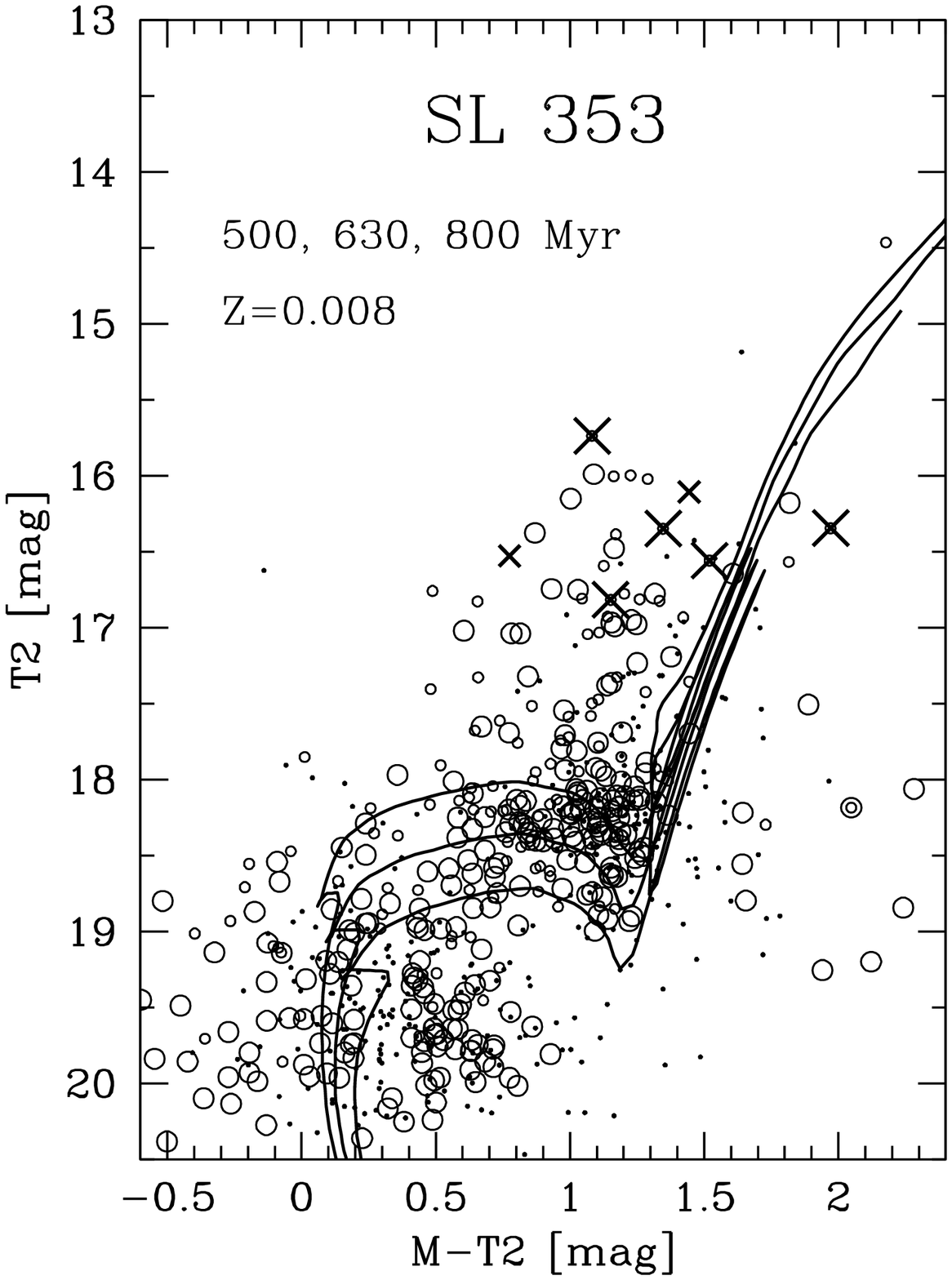}
\hfill
\includegraphics[width=8.8cm]{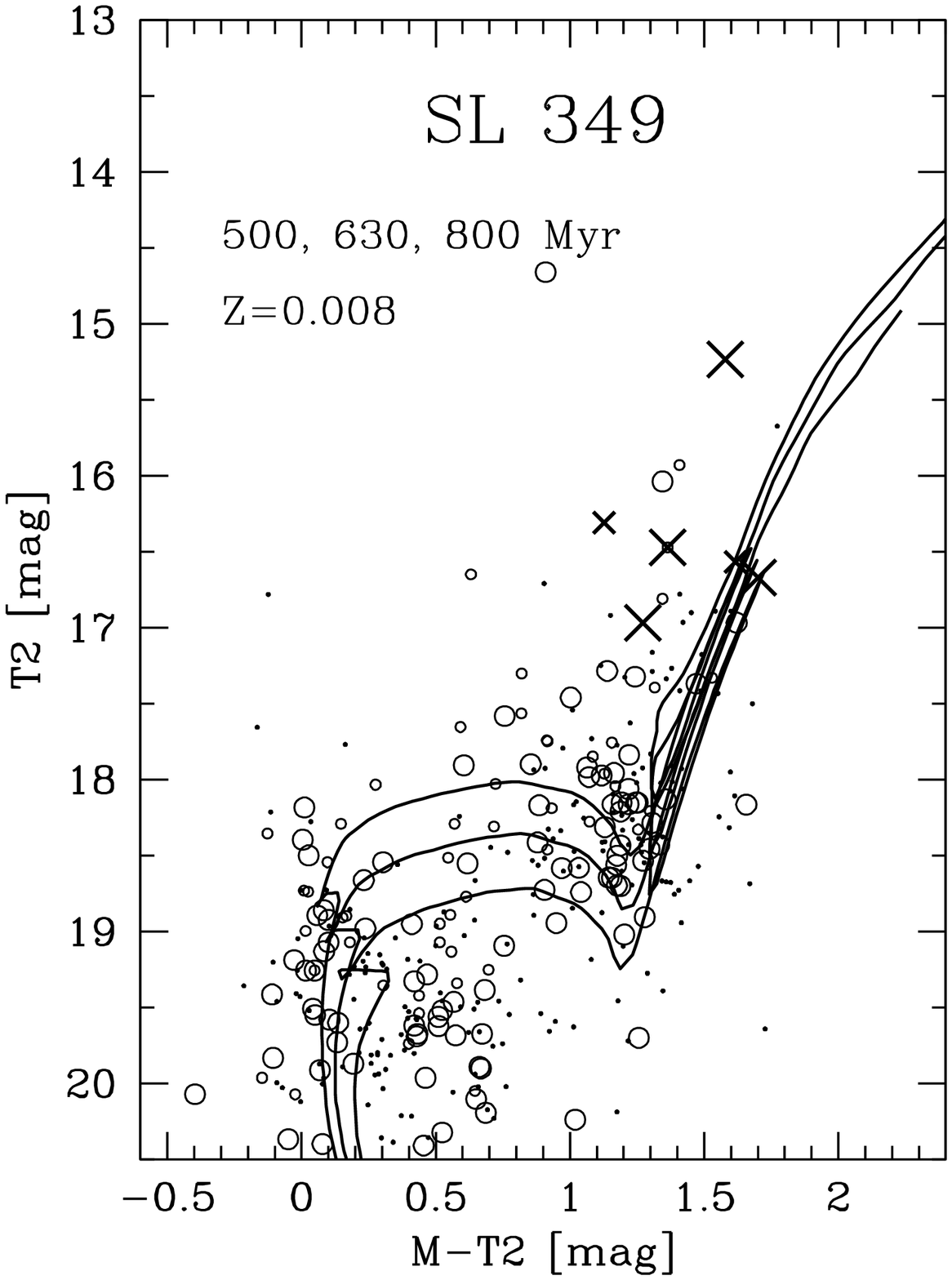}
}
\hfill
\begin{minipage}[t]{8.8cm} 
\resizebox{12cm}{!}
\centerline{
\includegraphics[width=8.8cm]{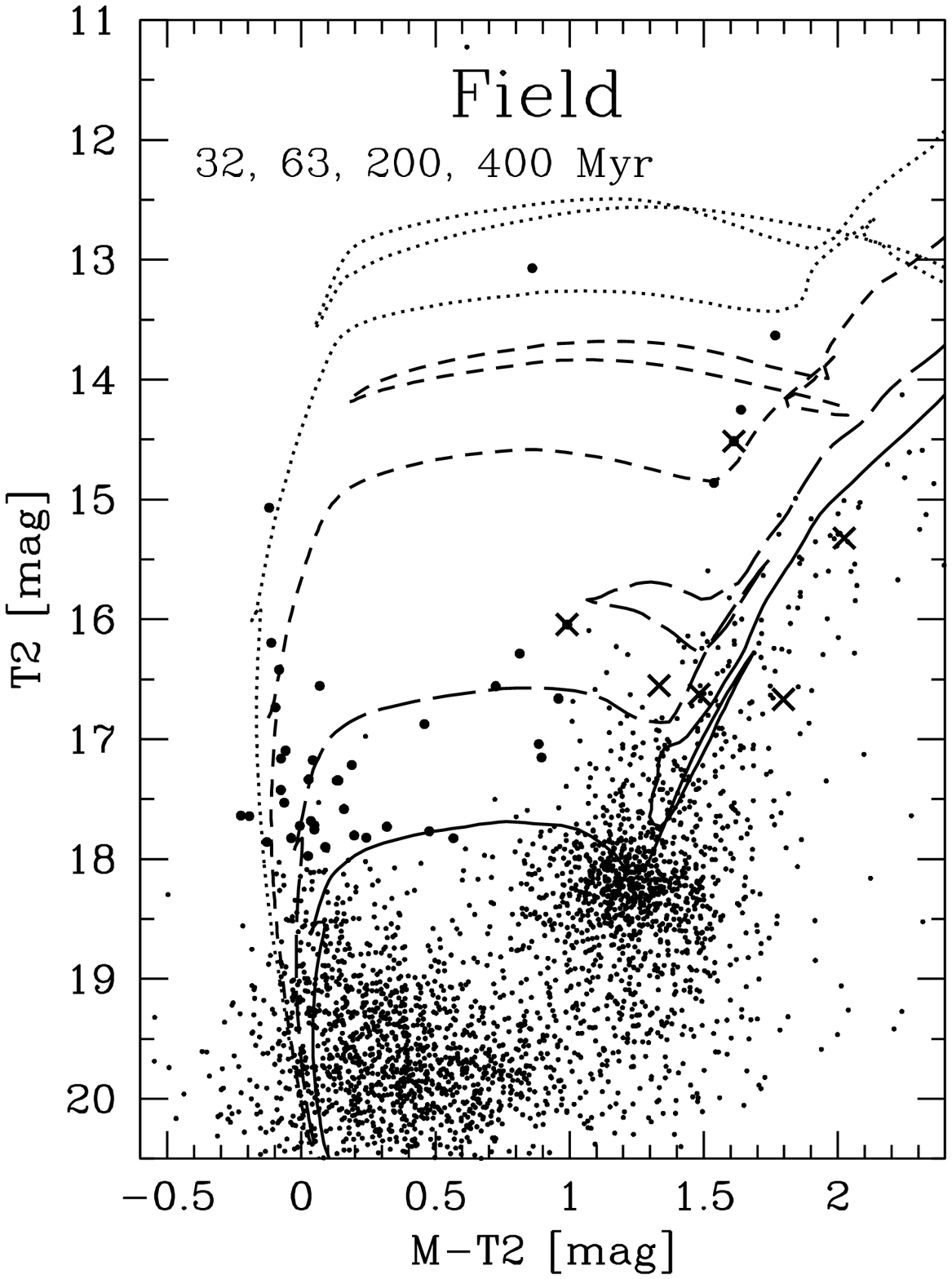}
}
\end{minipage} 
\hspace{0.95cm}
\begin{minipage}[t]{8cm}
\caption[]{\label{cmdsl353} Colour magnitude diagrams of the cluster SL\,353
  (upper left) and its companion SL\,349 (upper right). The data points
  represent all stars which are located in an annulus of 120 pixels
  corresponding to $40\farcs8$ around SL\,353, and of 100 pixels corresponding
  to $34\arcsec$ around SL\,349, respectively. The open circles represent
  stars that remain after statistical field star subtraction while the small
  dots represent field stars contaminating the cluster area. Small open
  circles denote stars lying inside a smaller annulus ($17\arcsec$ for
  SL\,353 and $13\farcs6$ for SL\,349) where crowding is most severe. The
  crosses mark stars for which also radial velocities could be derived; small
  crosses denote stars with velocities that are concordant with field
  velocities. The colour of the stars redwards of the main sequence may be
  caused by the crowding in the clusters' centres. Several possible isochrones
  are overplotted on the CMDs. Due to crowding a definitive age determination
  is difficult. We adopt an age of $550\pm100$ Myr for both clusters. Note the
  difference in age when fitting the main sequence turnoff or the red
  clump. The lower diagram shows the CMD of the surrounding field populations
  which comprise a mixture of ages. Stars with $T2\ge18$ mag and red
  giants are represented as smaller dots to keep the isochrones recognizable.
  See text for more details}
\vfill 
\end{minipage}
\end{figure*}

To derive the CMDs for each cluster we cut out a circular area
centred around the optical centre of each cluster. The radius was chosen to
be 100 pixels (corresponding to $34\arcsec$ or 8.3 pc) for SL\,349, and 120
pixels (corresponding to $40\farcs8$ or 9.9 pc) for SL\,353. An investigation
of the star density plots suggests that most cluster stars are located inside
the chosen areas. The cluster CMDs were compared with the CMD of the
surrounding field and a statistical field star subtraction was applied to
clear the cluster CMDs from contaminating field stars which might affect the
age determination. To take into account the different area sizes of the
regions covered by the clusters and the surrounding field we weighted the
field star subtraction with field sizes. 

We derived ages for the clusters by comparing the CMDs with isochrones which
are based on the stellar models of the Geneva group (Schaerer et
al. \cite{smms}). The isochrones were transformed to the Washington system by
Roberts \& Grebel (\cite{rg}).  

A distance modulus of 18.5 mag (Westerlund \cite{westerlund}) was adopted. We
assumed a metallicity of Z = 0.008 corresponding to $[Fe/H]\approx-0.3$ dex
which was found by various authors for the young field star population 
(e.g., Russell \& Bessell \cite{rb}, Luck \& Lambert \cite{ll}, Russell \&
Dopita \cite{rd}, Th\'{e}venin \& Jasniewicz \cite{tj}). 

Galactic field stars may contaminate our observed area. In order to assess
whether foreground stars might affect our age determination we compare our
CMDs with the number of galactic field stars towards the LMC estimated by
Ratnatunga \& Bahcall (\cite{raba}). In Table \ref{foreground} we present
their counts scaled to our field of view of $4\farcm3 \times 3\farcm5$. As
can be seen, the number of expected foreground stars is small, and we do not
expect an influence on our age determination.  

\begin{table}
\caption[]{\label{foreground}Number of foreground stars towards the LMC
  calculated from the data of Ratnatunga \& Bahcall (\cite{raba}), scaled to
  our field of view of $4\farcm3 \times 3\farcm5$}
\begin{tabular}{lccccc}
\hline
                    & \multicolumn{5}{c}{apparent visual magnitude range}\\
colour range        &13-15&15-17&17-19&19-21&21-23 \\
\hline
$B-V < 0.8$       & 0.6 & 1.3 & 1.4 & 2.9 & 2.7\\
$0.8 < B-V < 1.3$ & 0.2 & 1.2 & 2.4 & 1.9 & 3.3\\
$1.3 < B-V$       & 0.0 & 0.3 & 1.7 & 5.7 & 13.1\\
\hline
\end{tabular}
\end{table}

The CMDs of the two clusters and the surrounding field and the isochrone
fitting are described in more detail in the following.

{\it SL\,353:} The CMD of this cluster is shown in Fig. \ref{cmdsl353},
upper left plot. Open and filled data points represent all stars
which are located inside a radius of $40\farcs8$. After statistical field star
subtraction only the open circles remain. The CMD shows a wide blue main
sequence. The width of the main sequence is caused in part by photometric
errors (average seeing $1\farcs4$) and crowding. Also a pronounced red clump
of He-core burning stars and some red giants can be seen. Most red clump stars
remain after field star subtraction, an indication of the intermediate age of
this cluster. The ratio of red clump stars in the cluster and in the
surrounding field is approximately 3:1. The small open circles denote
the stars that can be found inside an annulus of $17\arcsec$ where crowding is
most severe. The width of the main sequence and the scatter of the red giants
is caused in part by this crowding. Overplotted on the CMD are several possible
isochrones. An isochrone resulting in an age of 500 Myr (upper line) gives a
good fit to the apparent main sequence turnoff at $T2\approx18.5$
mag. However, the evolved stars are not well represented by this
isochrone. Fitting the He-core burning giants results in higher ages of 630
Myr (middle line) or 800 Myr (lower line) but also underestimates the
luminosity of the main sequence turnoff. Since the fit to the main sequence
provides a lower age limit we adopt a mean age of 550$\pm$100 Myr for
SL\,353.     

Such a discrepancy between the main sequence turnoff and red giants
luminosities was noticed in the CMDs of various other LMC star clusters, two
prominent examples are NGC\,1866 and  NGC\,1850 (see Brocato et
al. \cite{bbcw}, Lattanzio et al. \cite{lvbc}, Vallenari et al. \cite{vafcom},
Brocato et al. \cite{bcp}). Unresolved binary stars which could increase the
luminosity of the main sequence turnoff were proposed as one
explanation. Other possibilities include blue stragglers and/or rotation
(Grebel et al. \cite{grebel}). Vallenari et al. (\cite{vafc}) point out that
no direct evidence for a significant population of binary stars could be
found, though the detection of close binary stars in the Magellanic Clouds is
difficult if not impossible. Lattanzio et al. (\cite{lvbc}) was able to solve
the problem by taking into account unresolved binaries as well as
semiconvection. Their simulated CMDs then gave a good representation of the
observed ones.  

{\it SL\,349:} Fig. \ref{cmdsl353}, upper right plot, presents the CMD of
SL\,349. Again, large and small open circles denote stars that remain after
field star subtraction. SL\,349 is the smaller one of the cluster pair, as can
also be seen from the sparse main sequence and red giant clump. The
small, open circles represent the stars located inside an inner, crowded annulus
of $13\farcs6$. Fitting an isochrone to the apparent main sequence turnoff
($T2\approx18.5$ mag) results in an age of 500 Myr (upper line). However, note
again the pronounced red clump that is present in the CMD. We find
approximately twice the amount of red clump stars in the cluster than in the
field, scaled to the same area. Fitting isochrones to the core-He-burning red
giants results in somewhat higher ages than isochrone fits that take into
account the apparent main sequence turnoff, namely 630 Myr (middle line) or
800 Myr (lower line). We adopt the same age of 550$\pm$100 Myr for SL\,349
that we already found for SL\,353.  

Our findings are in agreement with former age determinations:
Vallenari et al. (\cite{vbc}) found from CMDs a very similar age of
500$\pm$100 Myr for both clusters. 
Based on integrated colours, Bica et al. (\cite{bcdsp}) suggest that the
clusters are of somewhat older SWB type V (800 -- 2000 Myr).
Studying the stellar content of proposed binary clusters in the LMC, Kontizas
et al. (\cite{kkx}) found the older cluster pairs in their sample, among them
SL\,353 \& SL\,349, younger than 600 Myr, which agrees with our findings
and those  of Vallenari et al. (\cite{vbc}). 

{\it The surrounding field:} The CMD of the surrounding field is shown in the
lower diagram of Fig. \ref{cmdsl353}. A mixture of populations of different
ages can be found around the cluster pair. The CMD shows a bright blue main
sequence and a few red supergiants which represent the young population. The
intermediate age populations show up through the pronounced red clump of
He-core burning stars and the red giants. Distinct young populations cannot be
distinguished, but the overplotted isochrones are supported by corresponding
supergiants. Fainter main sequence stars and red giants are plotted with
smaller data points to keep the isochrones recognizable. The dotted isochrone
fits to the brightest main sequence stars and to some supergiants, resulting
in an age of 32 Myr. Also the 63 Myr isochrone (short dashed line) gives a
good representation of the few red supergiants. We see again a few bright
supergiants at $T2\approx16.5$ mag which can be fitted with a 200 Myr
isochrone (long dashed line). Between 200 Myr and 400 Myr (solid line) the
star density along the subgiant branch seems to be lower, indicating a
possible decrease in the field star formation rate, but increases again along
and below the 400 Myr isochrone (solid line). 

Since the main sequences in the CMDs show a large scatter, it is impossible to
derive a reddening via isochrone fitting. Thus, we adopt a reddening of
$E_{B-V}=0.1$ mag, corresponding to $E_{M-T2}=0.15$ mag (see Grebel \& Roberts
\cite{gr}, their Table 5, for the transformation of extinctions in different
filter systems), as suggested from the reddening maps of Schwering \& Israel
(\cite{si}).    

\section{Spectroscopic data and their reduction}
\label{spectroscopy}

In order to derive radial velocities, multi-object spectra (MOS) for 22 stars
in and around the star clusters SL\,535 \& SL\,349 were obtained on February
09, 1995, using EMMI RILD at the ESO/NTT at La Silla. A TEK $2048\times2048$
chip (ESO \#36) was used with a pixel scale of $0\farcs27$. The resulting
field of view is $9\farcm2\times8\farcm6$. Slit masks are prepared at the NTT
during observation, the punch field covers $5\arcmin\times8\arcmin$. Only 19
stars out of 22 for which spectra were obtained are located inside the smaller
field of view of the photometric data ($4\farcm3 \times 3\farcm5$). They are
marked in Fig. \ref{sl353ps}. The spectra were taken with grism \#6 which
covers a wavelength range of 6000 to 8300 {\AA} when the slits are centred on
the CCD chip, and provides a dispersion of 1.2 {\AA} per pixel.  
The brightest stars in both clusters are red giants (see Sect. \ref{ages}), so
we chose to obtain spectra of red giants at the Ca\,II triplet at $\lambda =$
8498~{\AA}, 8662~{\AA} and 8542~{\AA} (see e.g. Olszewski et
al. \cite{ossh}). The wavelength coverage of the MOS depends on the position
of the slits with respect to the centre of the CCD. In order to reach the
Ca\,II lines the X-position of the slits was shifted about $\approx2\farcm5$
or $\approx530$ pixels resulting in a wavelength coverage of 6500 to 9000
{\AA}. The slit width was $\approx1\arcsec$ (corresponding to 4 pixels). In
the same night the radial velocity standard stars HD\,101266, HD\,111417, and
HD\,120223 were observed using the same slit mask. An observing log is given
in Table \ref{speclog}. The velocities of the standard stars were taken from
Evans (\cite{evans}).    

Data reduction was carried out using IRAF's {\tt twodspec} and
{\tt onedspec} packages. We mainly followed the procedure for multislit
spectroscopic reductions described in Ellingson (\cite{ellingson}). Wavelength
calibration was performed using He-Ar spectra obtained in the same night with
the same slit mask. 

\begin{table}
\caption[]{\label{speclog}Observing log of the spectroscopic data}
\begin{tabular}{rcccc}
\hline
Object             & $\alpha_{2000}$  & $\delta_{2000}$ & Exp.time & velocity\\
 &[$^{\mbox{h\,m\,s}}$]&[$^{\circ}\,\arcmin\,\arcsec$]&[s]&[km\,$\mbox{s}^{-1}$]\\\hline 
SL\,353/349        & 05 16 00         & $-68$ 56 00.0 & 3600   & \\
HD\,101266         & 11 38 50.73      & $-45$ 21 45.3 & 30     & $+20.6$ \\
HD\,111417         & 12 49 31.82      & $-45$ 49 32.8 & 30     & $-16.0$ \\
HD\,120223         & 13 49 06.56      & $-43$ 44 00.1 & 30     & $-24.1$ \\
\hline
\end{tabular}
\end{table}

\section{Radial velocities}
\label{radvel}

\begin{figure*}
\centerline{
\includegraphics[width=\hsize]{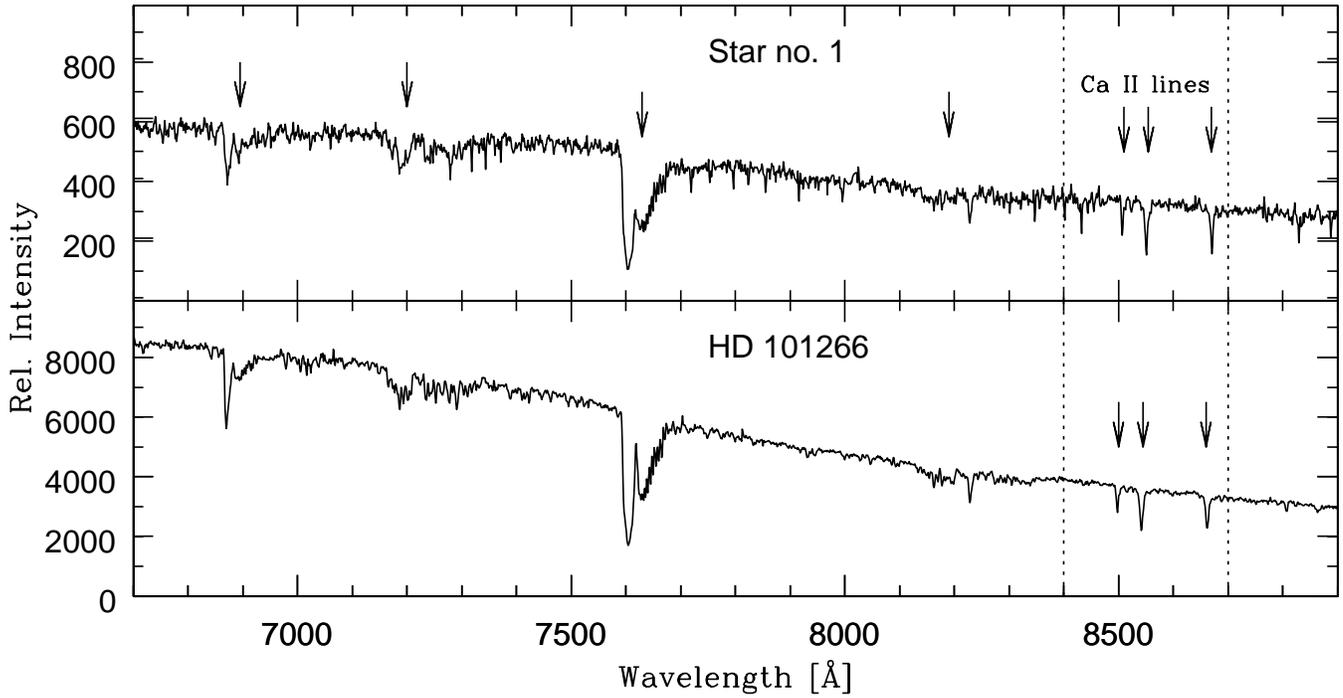}
}
\caption[]{\label{rvspec} The upper panel shows the spectrum of the Star no. 1
while the lower panel shows the spectrum of the radial velocity standard star
HD\,101266. Strong night sky absorption lines are marked with arrows (only in
the upper panel): The two sharp peaks at $\lambda =$ 6867 {\AA} and $\lambda
=$ 7594 {\AA} denote to the $\mbox{O}_{2}$ B-band and $\mbox{O}_{2}$ A-band,
the other two more extended bands starting at $\lambda =$ 7186 and 8164 {\AA},
respectively, are caused by atmospheric water vapor. For the crosscorrelation
only the wavelength range between the dotted lines was used. The Ca\,II
triplet can be clearly seen in both spectra }
\end{figure*}

\begin{figure*}
\centerline{
\includegraphics[height=22cm]{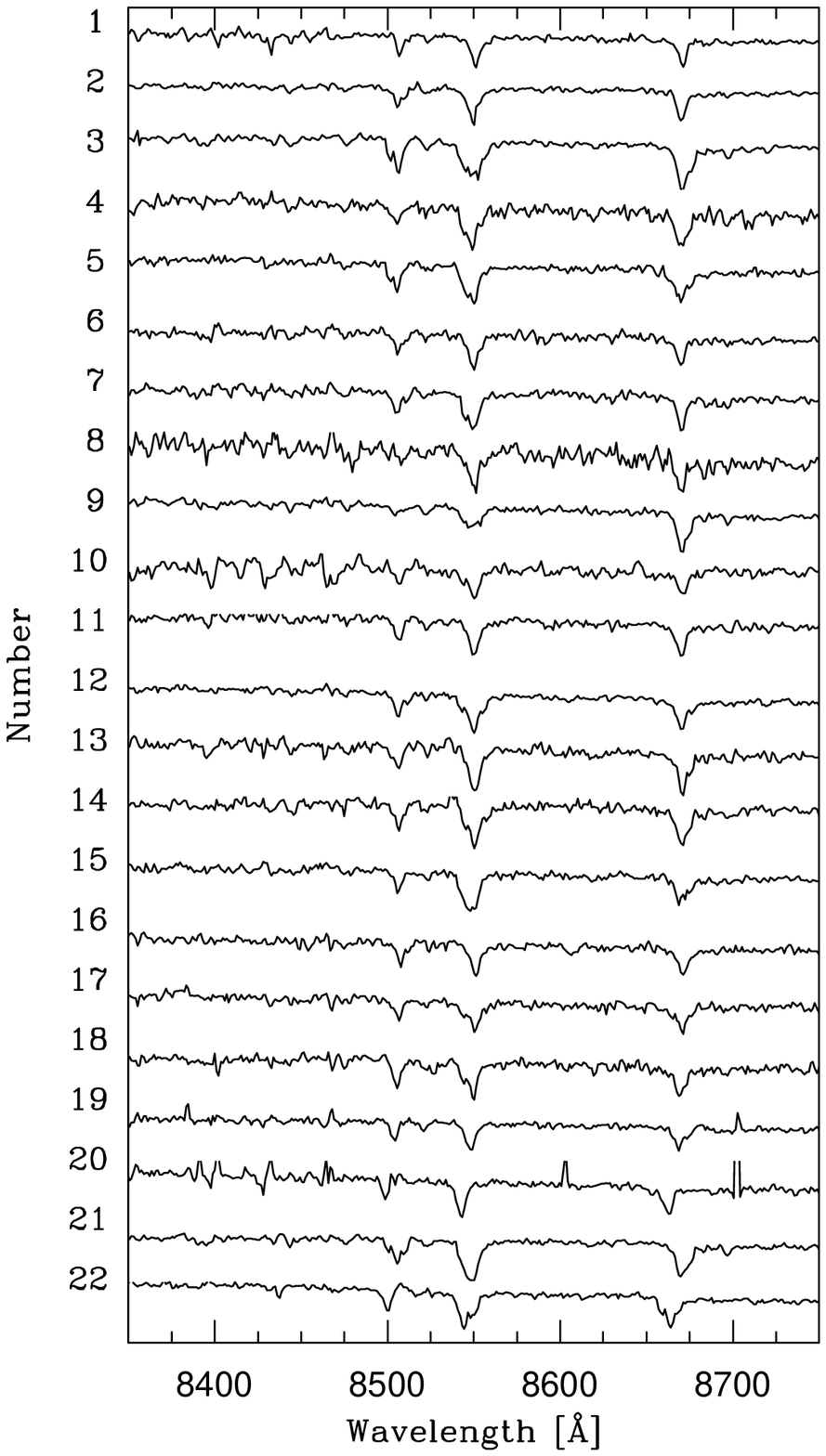}
}
\caption[]{\label{spectra3} Spectra of the stars no. 1 to 22. Only the
  wavelength range of 8350 {\AA} to 8750 {\AA} which includes the Ca\,II lines
  and which was used for the crosscorrelation is plotted. The Ca\,II triplet
  rest wavelenghts are at $\lambda =$ 8498~{\AA}, 8662~{\AA} and
  8542~{\AA}. The Ca\,II lines for the stars no. 20 and 22 are less displaced
  compared to the other spectra which identifies them as galactic foreground
  stars. See text for more details} 
\end{figure*}

We derived the radial velocities for the 22 stars by cross-cor\-re\-lating each
spectrum with the spectra of the three standard stars listed in Table
\ref{speclog}. An example is given in Fig. \ref{rvspec}. The upper panel shows
the spectrum of the Star no. 1 (which lies outside of the useful area of our
imaging data) while the lower panel shows the spectrum of the radial velocity
standard star HD\,101266. The whole wavelength range covered by the grism is
plotted. Strong atmospheric absorption lines are visible in both spectra. The
sharp peak at $\lambda =$ 6867 {\AA} denotes the $\mbox{O}_{2}$ B-band, and
the even stronger double peak at $\lambda =$ 7594 {\AA} belongs to the
$\mbox{O}_{2}$ A-band (Turnshek et al. \cite{ttcb}). Two
$\mbox{H}_{2}\mbox{O}$-bands are visible, starting at $\lambda =$ 7186 and
8164 {\AA}, respectively.   

Using the IRAF task {\tt fxcorr} in the package {\tt rv} we applied a
crosscorrelation between each programme and standard star spectrum. For this
purpose we only need the Ca\,II lines. In Fig. \ref{rvspec} the region used
for the crosscorrelation is marked with dotted lines. Table \ref{velocity} 
gives an overview of the radial velocities for all observed stars. The
velocities listed are the error weighted mean of the three velocities derived
from the crosscorrelation between the programme and the three standard stars. 
 
In Fig. \ref{spectra3} the spectra of all 22 programme
stars are shown. We only plotted a wavelength range of 8350~{\AA} to
8750~{\AA} which includes the Ca\,II lines from which we derived the
velocities. The two strongest lines of the Ca\,II triplet can be clearly seen
in all spectra at $\approx 8670$ {\AA} and $\approx 8550$ {\AA}. The third and
weakest line appears at $\approx 8500$ {\AA}. 

Star no. 8 is the faintest star of our sample and thus shows the noisiest
spectrum (see Fig. \ref{spectra3}), which is also responsible for the large
error of the velocity (see Table \ref{velocity}). Also star no. 10 shows a
somewhat noisy spectrum and has a higher $\sigma$. Stars no. 20 and 22 must be
foreground stars since their velocities are small (14.4 km\,$\mbox{s}^{-1}$
and 84.3 km\,$\mbox{s}^{-1}$, respectively). The minor displacement of the
Ca\,II lines can be clearly seen in Fig.~\ref{spectra3}. All other  stars have
velocities which indicate that they are LMC members (approximately 200 -- 320
km\,$\mbox{s}^{-1}$, see e.g. Pr\'{e}vot et al. \cite{prevot}, or Westerlund
\cite{westerlund}). Stars no. 1 and 
16 have velocities higher than 300 km\,$\mbox{s}^{-1}$. These two stars might
be field stars having such a high velocity, or they may be variable or binary
stars (see e.g. Fischer et al. \cite{fwm}). 

The programme stars are located in the field as well as in both clusters and
in between the two clusters. It is striking that most stars located inside the
cluster areas have higher velocities while (almost) all field stars show lower
velocities (see Table \ref{velocity}). The components of our proposed binary
cluster have very similar mean radial velocities of $\approx274\pm10$
km\,$\mbox{s}^{-1}$ for SL\,349 (stars no. 6, 8, 9, 10) and $\approx279\pm4$
km\,$\mbox{s}^{-1}$ for SL\,353 (stars no. 11, 12, 13, 14, 17). The mean
velocity for both clusters is $\approx277\pm7$ km\,$\mbox{s}^{-1}$. In
contrast the investigated field stars have velocities of $\approx240\pm19$
km\,$\mbox{s}^{-1}$. The errors given for these mean velocities are the
standard deviations. 

The expected internal velocity dispersion of a star cluster is of
the order of only few km\,$\mbox{s}^{-1}$ (see, e.g., Westerlund
\cite{westerlund}). However, since the mean error of one single velocity
measurement is $\approx6.6$ km\,$\mbox{s}^{-1}$ we do not state about the
internal velocity dispersion of SL\,349 or SL\,353.           

\begin{table}
\caption[]{\label{velocity} Radial velocities of all 22 programme stars}
\begin{tabular}{ccccc}
\hline
star no.    & velocity          & $T2$  & $M-T2$ & comment\\
            & [km\,$\mbox{s}^{-1}$]          & [mag] & [mag]  &        \\
\hline\hline
1           & 313.7$\pm$3.1     & ---   & ---  & variable/binary?\\
2           & 265.5$\pm$2.5     & ---   & ---  & field star\\
3           & 268.5$\pm$3.7     & 14.52 & 1.61 & field star\\
4           & 231.1$\pm$2.4     & 16.62 & 1.48 & field star\\
5           & 232.7$\pm$2.9     & 16.31 & 1.12 & field star\\
6           & 275.2$\pm$2.6     & 16.67 & 1.70 & cluster star\\
7           & 248.7$\pm$3.2     & 16.57 & 1.62 & field star\\
8           & 283.9$\pm$8.2     & 16.97 & 1.27 & cluster star\\
9           & 260.2$\pm$5.0     & 15.24 & 1.58 & cluster star\\
10          & 277.6$\pm$9.8     & 16.47 & 1.36 & cluster star\\
11          & 279.4$\pm$3.4     & 16.56 & 1.52 & cluster star\\
12          & 275.3$\pm$2.7     & 15.74 & 1.08 & cluster star\\
13          & 284.0$\pm$2.5     & 16.35 & 1.35 & cluster star\\
14          & 280.4$\pm$3.5     & 16.35 & 1.97 & cluster star\\
15          & 235.1$\pm$4.2     & 16.11 & 1.44 & field star\\
16          & 318.5$\pm$2.7     & 16.53 & 0.78 & variable/binary?\\
17          & 275.2$\pm$3.6     & 16.82 & 1.15 & cluster star\\
18          & 238.3$\pm$3.6     & 16.67 & 1.80 & field star\\
19          & 206.9$\pm$2.7     & 16.56 & 1.33 & field star\\
20          &  14.4$\pm$3.2     & 16.05 & 1.00 & foregr. star\\
21          & 229.0$\pm$3.3     & 15.32 & 2.02 & field star\\
22          &  84.3$\pm$3.4     &  ---  & ---  & foregr. star\\
\hline
\end{tabular}
\end{table}

\section{N-body simulations}
\label{nbody}

\begin{figure*}
\centerline{
\includegraphics[height=22cm]{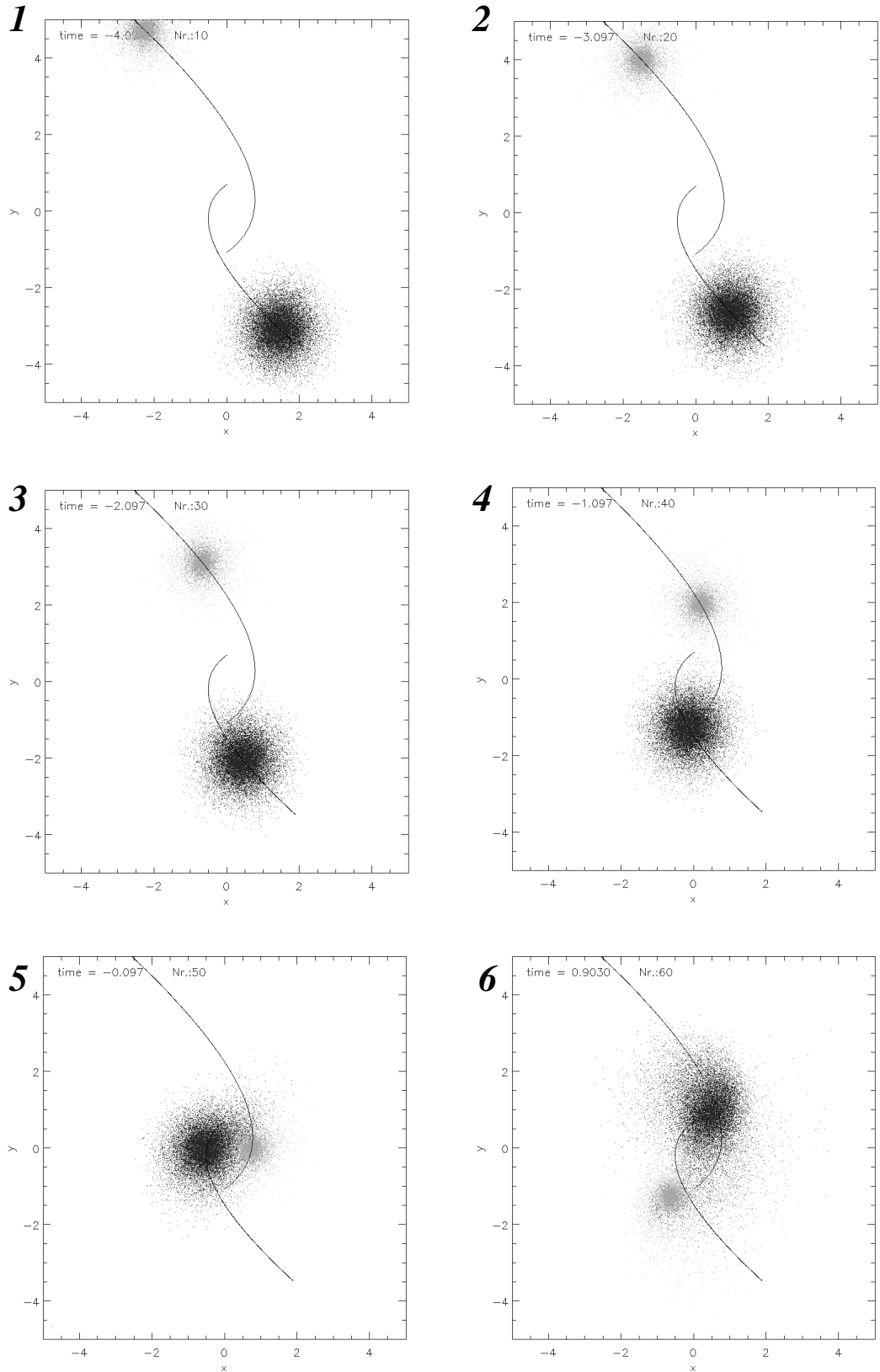}
}
\caption[]{\label{nbodysim} N-body simulation of a close encounter of two star
  clusters which resemble our observed cluster pair. The two clusters approach
  each other on elliptical orbits under the influence of their gravitational
  forces. The tidal field of the parent galaxy is not considered in this
  simulation} 
\end{figure*}

\begin{figure}
\centerline{
\includegraphics[width=\hsize]{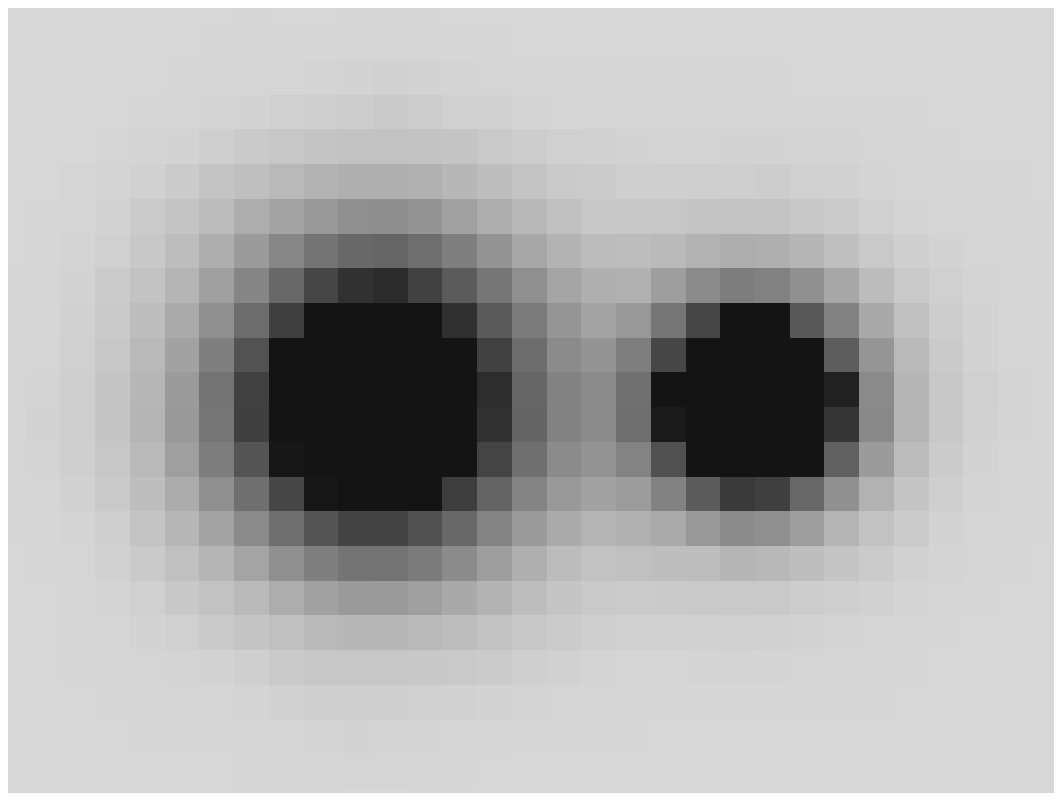}
}
\caption[]{\label{king1} Density plot of two artificial star clusters without
  any interaction. The centre-to-centre separation between the clusters is the
  same as the projected distance between the observed pair SL\,353 \& SL\,349
  (18.1 pc). The star density between the clusters, where the density profiles
  are overlapping, is enhanced} 
\vspace{0.4cm}
\centerline{
\includegraphics[width=\hsize]{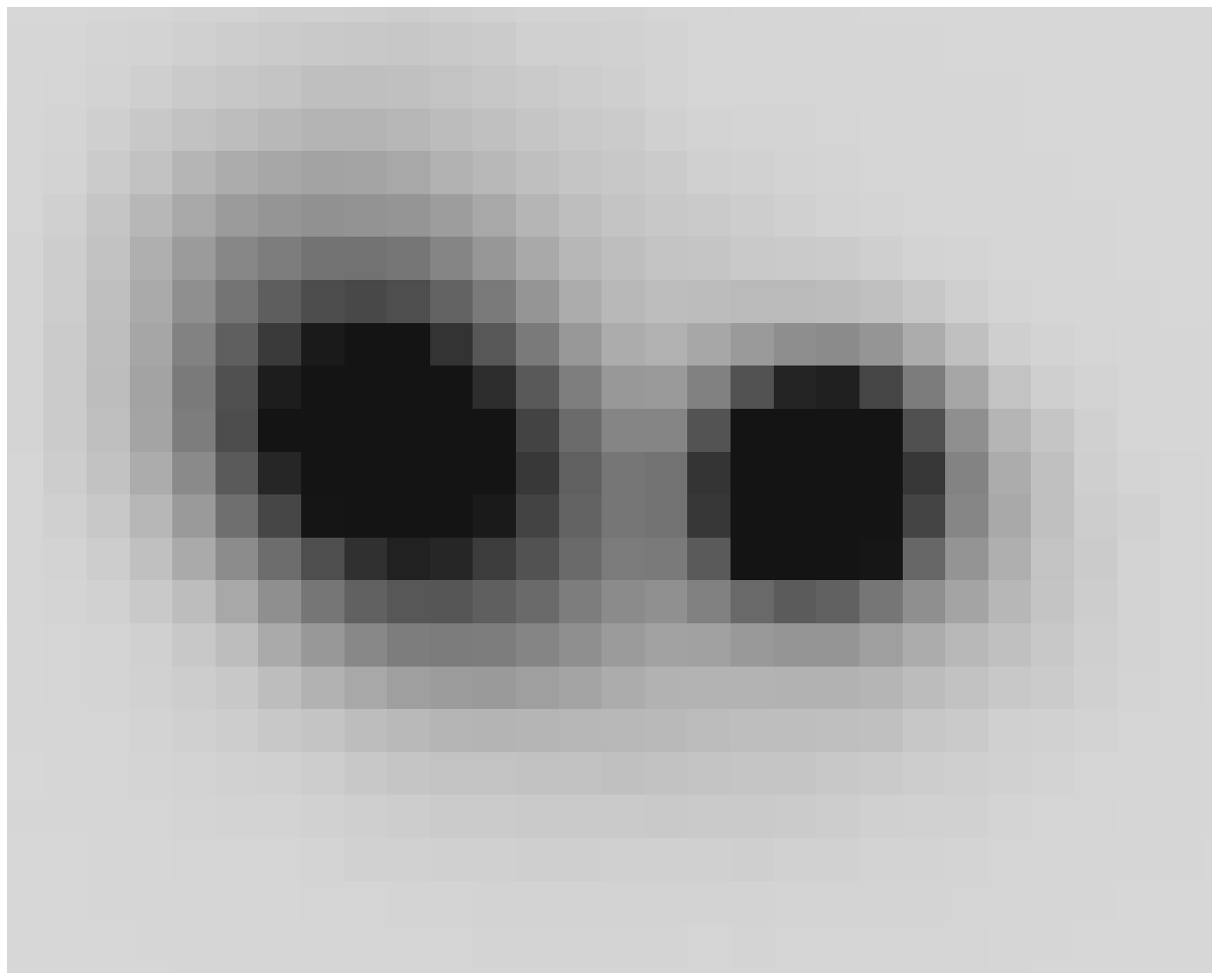}
}
\caption[]{\label{king2} Isopleth of the two artificial clusters with
  interaction, the distance between the two clusters is the same as in
  Fig. \ref{king1}. The region between the clusters shows enhanced star
  density. The larger cluster shows a distortion while the smaller but more
  compact one remains mainly undisturbed in the isopleth}
\end{figure}

Whether two clusters are interacting or not is hard to decide on the base of
imaging data. According to de\,Oliveira et al. (\cite{dodb}) encounters
of star clusters can be traced using isodensity maps. However, the field in
which the clusters are embedded may likely influence the isopleths, and the
distinction between field and cluster stars is a difficult task (Leon et
al. \cite{lbv}).  
   
Assuming a Kroupa IMF (Kroupa et al. \cite{kroupa}) we estimate the total 
masses, the number of stars, and the half-mass and tidal radii of the
star clusters SL\,353 and SL\,349. The observational parameters can be
found in Table \ref{kingpars}.  

\begin{table}
\caption[]{\label{kingpars}Observational parameters used to create artificial
  star clusters}
\begin{tabular}{lcccc}
\hline
cluster & total mass    & number   & halfmass radius & tidal radius\\
        & [$M_{\odot}$] & of stars & [pc]            & [pc]  \\ \hline
SL\,353 & 7300          & 20100    & 6.6             & 23.0  \\ 
SL\,349 & 4800          & 13100    & 3.3             & 20.6  \\ \hline 
\end{tabular}
\end{table} 

Based on a King model and the parameters listed in Table \ref{kingpars} two
artificial star clusters were created which resemble the observed
ones. Fig. \ref{king1} shows the density plot of the two artificial clusters
placed at the same distance as the projected separation between the cluster 
pair SL\,353 \& SL\,349 (18.1 pc). The star density between the clusters,
where the density profiles are overlapping, is enhanced.   

In order to get an idea whether an interacting cluster pair could be discerned
from a mere overlap by chance in the line-of-sight, we performed N-body
simulations using the GRAPE3a special purpose board in Kiel. The equations of
motion were integrated using the standard leapfrog scheme with a constant time
step of 0.15 Myr. Additionally, we allowed for a Plummer softening with a
softening length of 0.1 pc. Since we assume that the cluster pair SL\,353 \&
SL\,349 is a 'real' pair we chose a parabolic orbit as a limiting case to
unbound pairs. In case of a 'true' gravitationally bound cluster pair, one
would expect elliptic orbits and, thus, a series of repetitive encounters
which should result in even stronger signatures of an interaction or
eventually already in a merger. At the beginning we started with a well
detached system at a separation of 100 pc. The minimum distance (reached at
t=0) was set to 12 pc. The simulation was stopped at separation of 18.1 pc,
i.e. 13.4 Myr after closest approch. For simplicity, the tidal field of the 
parent galaxy is not considered in this simulation. 
Fig. \ref{nbodysim} shows a face-on view of both clusters.

The isopleth of the interacting clusters, again at a distance of 18.1 pc, is
presented in Fig. \ref{king2}. Again, a region of enhanced star density
connects both clusters. Furthermore the larger cluster shows a distortion
while the smaller but more compact one remains mainly undisturbed in the
isopleth. The density plot of the observed cluster pair (Fig. \ref{sl353dens})
shows a similar, northeast to south distortion around the more massive
cluster SL\,353. However, the observed northern distortion and the distortion
of the smaller cluster SL\,349 are not seen in Fig. \ref{king2}.

We want to emphasize that the parameters used to create the artificial
clusters are based on our observations, but are only estimates and should not
be taken as strict values. The simulations are meant for qualitative comparison
and a more detailed simulation is barely possible since the parameter space
cannot be further restricted on the base of our data.  

\section{Summary and conclusions}
\label{summary}

Fitting isochrones based on the Geneva models (Schaerer et al. \cite{smms}) to
the CMDs, we find that the components of the double cluster SL\,353 \& SL\.349
are coeval within the accuracy of our data, and we derive an age of
550$\pm$100 Myr for both clusters. 

The clusters are sufficiently old that the Ca\,II triplet visible in the
spectra of red giants could be used to derive radial velocities. 22 stars in
and around the two star clusters were investigated. Most stars located inside
the cluster areas show similar velocities of $\approx277\pm7$
km\,$\mbox{s}^{-1}$ whereas almost all field stars show lower velocities of
$\approx240\pm19$ km\,$\mbox{s}^{-1}$. Two foreground stars were identified
through their velocities which are too low to belong to the LMC population (14
km\,$\mbox{s}^{-1}$ and 84 km\,$\mbox{s}^{-1}$, respectively).   

Both components of the binary cluster candidate are of the same age and,
furthermore, have very similar mean radial velocities of $\approx274\pm10$
km\,$\mbox{s}^{-1}$ for SL\,349 and $\approx279\pm4$ km\,$\mbox{s}^{-1}$ for
SL\,353. These findings support that both clusters may have formed
at similar times and from the same GMC. In this sense they may constitute a
true binary cluster.  

We investigate the stellar densities in and around the binary cluster SL\,353
\& SL\,349 and find an enhanced density between the two clusters. The smaller
cluster SL\,349 shows a distortion towards SL\,353, and similar distortions
can be seen around the more massive cluster.  

Vallenari et al. (\cite{vbc}) found a distortion of the isophotal contours,
similar to our Fig. \ref{sl353dens}, and a twisting of the isopleths which
they regard as a sign of interaction and physical connection between the two
clusters. It is remarkable that the age of this binary cluster is higher than
the theoretical survival time of few $10^{7}$ yrs for physically connected
cluster pairs suggested by Bhatia (\cite{bhatia}). Leon et al. (\cite{lbv})
suggest that SL\,349 \& SL\,353 are part of a larger star cluster group in
which interacting binary clusters are formed later via tidal capture, thus
explaining an age larger than the theoretical survival time. However, another
explanation could be that the survival time might be larger in the LMC bar
where the tidal field might be weaker, as proposed by Elson et
al. (\cite{elson}, their Fig.\,13). 

Whether the distortion of the clusters is indeed a sign of possible
interaction cannot be decided on the basis of our imaging data alone.       

Based on observational parameters we created two artificial
star clusters. The isopleths of the non-interacting pair as well as of
a simulated interacting pair of star clusters were compared with the 
observed density plot. The density plot of the artificial, interacting pair
shows a distortion of the more massive cluster which can also be found in the
observed isopleth, however, the observed distortion of the smaller cluster
SL\,349 is not seen in the artificial density plot. It seems likely that this
cluster pair shows signs of interaction, however, this does not necessarily
imply that both clusters are gravitationally bound. 

\acknowledgements

We would like to thank Prof. H. Els\"{a}sser for allocating time
at the MPIA 2.2m-telescope at La Silla during which our imaging data were
obtained, Hardo M\"{u}ller for software advice, Klaas
S. de\,Boer and J\"{o}rg Sanner for a critical reading of the manuscript. AD
thanks Lick Observatory for their hospitality, where part of this work was
done. This work was supported by a graduate fellowship of the German Research
Foundation (Deutsche Forschungsgemeinschaft -- DFG) for AD through the
Graduiertenkolleg `The Magellanic System and Other Dwarf Galaxies' (GRK
118/2-96).   
EKG gratefully acknowledges support by NASA through grant HF-01108.01-98A from
the Space Telescope Science Institute, which is operated by the Association of
Universities for Research in Astronomy, Inc., under NASA contract NAS5-26555.
 
The simulations were carried out on the GRAPE3af special purpose computer in
Kiel (DFG Sp345/5). 

This research has made use of NASA's Astrophysics Data System Abstract
Service and of the SIMBAD database operated at CDS, Strasbourg, France.


\begin{thebibliography}{}
\bibitem[1990]{bhatia}
Bhatia R. K., 1990, PASJ 42, 757
\bibitem[1988]{bm}
Bhatia R. K., McGillivray H. T., 1988, A\&A 203, L5
\bibitem[1988]{bh}
Bhatia R. K., Hatzidimtriou D., 1988, MNRAS 230, 215
\bibitem[1991]{brht}
Bhatia R. K., Read M. A., Hatzidimtriou D., Tritton S., 1991, A\&AS 87, 335
\bibitem[1996]{bcdsp}
Bica E., Claria J. J., Dottori H., Santos J. F. C. Jr., Piatti A. E., 1996,
ApJS 102, 57
\bibitem[1989]{bbcw}
Brocato E., Buonanno R., Castellani V., Walker A. R., 1989, ApJS 71, 25
\bibitem[1994]{bcp}
Brocato E., Castellani V., Piersimoni A. M., 1994, A\&A 290, 59
\bibitem[1998]{dodb}
de\,Oliveira M.R., Dottori H., Bica E., 1998, MNRAS 295, 921
\bibitem[2000]{obd}
de\,Oliveira M.R., Bica E., Dottori H., 2000, MNRAS 311, 589 
\bibitem[1998]{dg}
Dieball A., Grebel E., 1998, A\&A 339, 773
\bibitem[1998]{ee}
Efremov Y., Elmegreen B., 1998, MNRAS 299, 588 
\bibitem[1997]{epth}
Ehlerov\'a S., Palou\v s J., Theis C., Hensler G., 1997, A\&A 328, 12
\bibitem[1989]{ellingson}
Ellingson E., 1989, A User's Guide to Multislit Spectroscopic Reductions with
IRAF, ftp://iraf.noao.edu/iraf/docs/
\bibitem[1999]{eepz}
Elmegreen B. G., Efremov Y. N., Pudritz R. E., Zinnecker H., 1999, to be
published in Protostars and Planets IV, eds. V. G. Mannings, A. P. Boss,
S. S. Russell, astro-ph/9903136  
\bibitem[1987]{elson}
Elson R. A. W., Fall M., Freeman K.C., 1987, ApJ 323, 54
\bibitem[1967]{evans}
Evans D. S., 1967, in IAU Symp. No. 30, eds. Batten A. H. and Heard J. F.,
p. 57   
\bibitem[1993]{fwm}
Fischer P., Welch D. L., Mateo M., 1993, AJ 105, 938
\bibitem[1997]{fk}
Fujimoto M., Kumai Y., 1997, AJ 113, 249
\bibitem[1990]{geisler}
Geisler D., 1990, PASP 102, 344
\bibitem[1995]{gr}
Grebel E., Roberts W. J., 1995, A\&As 109, 293
\bibitem[1996]{grebel}
Grebel E.K., Roberts W.J., Brandner W., 1996, A\&A 311, 470
\bibitem[1990]{hb}
Hatzidimitriou D., Bhatia R. K., 1990, A\&A 230, 11
\bibitem[1989]{kkx}
Kontizas E., Kontizas M., Xiradaki E., 1989, Ap\&SS 156, 81
\bibitem[1993]{kroupa}
Kroupa P., Tout C. A., Gilmore G., 1993, MNRAS 262, 545
\bibitem[1991]{lvbc}
Lattanzio J. C., Vallenari A., Bertelli G., Chiosi C., 1991, A\&A 250, 340
\bibitem[1999]{lbv}
Leon S., Bergond G., Vallenari A., 1999, A\&A 344, 450
\bibitem[1992]{ll}
Luck R. E., Lambert D. L., 1992, ApJS 79, 303
\bibitem[1991]{ossh}
Olszewski E. W., Schommer R. A., Suntzeff N. B., Harris H. C., 1991, AJ 101, 
515
\bibitem[1975]{page}
Page T., 1975, in Stars \& Stellar Systems, Vol. 9, p. 541, University of 
Chicago Press, Chicago
\bibitem[1985]{prevot}
Pr\'{e}vot L., Andersen J., Ardeberg A. et al., 1985, A\&AS 62, 23
\bibitem[1985]{raba}
Ratnatunga K., Bahcall J., 1985, ApJS 59, 63
\bibitem[1994]{rg}
Roberts W. J., Grebel E. K., 1994, American Astronomical Society Meeting
\#185, 104.02 
\bibitem[1989]{rb}
Russell S. C., Bessell M. S., 1989, ApJS 70, 865
\bibitem[1992]{rd}
Russell S. C., Dopita M. A., 1992, ApJ 384, 508
\bibitem[1993]{smms}
Schaerer D., Meynet G., Maeder A., Schaller G., 1993, A\&AS 98, 523
\bibitem[1991]{si}
Schwering  P. B. W., Israel F. P., 1991, A\&A 246, 231
\bibitem[1980]{swb}
Searle L., Wilkinson A., Bagnuolo W., 1980, ApJ 239, 803 
\bibitem[1991]{stetson}
Stetson P. B., 1991, 3rd ESO/ST-ECF Garching - Data Analysis Workshop, eds.
Grosb{\o}l P. J., Warmels R. H., p. 187
\bibitem[1999]{ss}
Subramaniam A., Sagar R., 1999, AJ 117, 937
\bibitem[1989]{sm}
Sugimoto D., Makino D., 1989, PASJ 41, 991
\bibitem[1991]{surdin}
Surdin V. G., 1991, Ap\&SS 183, 129
\bibitem[1998]{theis}
Theis C., 1998, in "Dynamics of Galaxies and Galactic Nuclei",
Proc. Ser. I.T.A., Vol. 2, eds. W. Duschl \& Ch. Einsel, p. 223
\bibitem[1997]{teph}
Theis C., Ehlerov\'a S., Palou\v s J., Hensler G., 1997, Proceedings of IAU
Coll. 166 "The Local Bubble and Beyond", Garching, p. 409 
\bibitem[1992]{tj}
Th\'{e}venin F. , Jasniewicz G., 1992, A\&A 266, 85
\bibitem[1985]{ttcb}
Turnshek D. E., Turnshek D. A., Craine E. R., Boeshaar P. C., 1985, Astronmy
and Astrophysics Series Vol. 1 
\bibitem[1994a]{vafc}
Vallenari A., Aparicio A., Fagotto F., Chiosi C., 1994a, A\&A 284, 424
\bibitem[1994b]{vafcom}
Vallenari A., Aparicio A., Fagotto F., et al., 1994b, A\&A 284, 447
\bibitem[1998]{vbc}
Vallenari A., Bettoni D., Chiosi C., 1998, A\&A 331, 506
\bibitem[1996]{vdbergh}
van den Bergh S., 1996, ApJ 471, L31
\bibitem[1997]{westerlund}
Westerlund B.E., 1997, `The Magellanic Clouds', Cambridge University Press,
Cambridge, UK 

\end{thebibliography}
\end{document}